**An introduction to statistical models used to characterize species-habitat associations with animal movement data**

Katie R.N. Florko[1,*], Ron R. Togunov[2,3,4], Rowenna Gryba[1,5,6], Evan Sidrow[5], Steven H. Ferguson[7,8], David J. Yurkowski[7,8], Marie Auger-Méthé[1,5]

[1]Institute for the Oceans and Fisheries, University of British Columbia, Vancouver BC, Canada

[2]Department of Zoology, University of British Columbia, Vancouver BC, Canada

[3]Department of Mathematical Sciences, Norwegian University of Science and Technology, Trondheim, Norway

[4]Centre for Biodiversity Dynamics, Norwegian University of Science and Technology, Trondheim, Norway

[5]Department of Statistics, University of British Columbia, Vancouver BC, Canada

[6]Department of Geography, University of British Columbia, Vancouver BC, Canada

[7]Fisheries and Oceans Canada, Freshwater Institute, Winnipeg MB, Canada

[8]Department of Biological Sciences, University of Manitoba, Winnipeg MB, Canada

***Corresponding author**:

Katie R.N. Florko

University of British Columbia

2202 Main Mall

Vancouver BC V6T 1Z4

Phone: 1-236-268-4471

Email: katieflorko@gmail.com

## Abstract

Understanding species-habitat associations is fundamental to ecological sciences and for species conservation. Consequently, various statistical approaches have been designed to infer species-habitat associations. Due to their conceptual and mathematical differences, these methods can yield contrasting results. In this paper, we describe and compare commonly used statistical models that relate animal movement data to environmental data. Specifically, we examined selection functions which include resource selection function (RSF) and step-selection function (SSF), as well as hidden Markov models (HMMs) and related methods such as state-space models. We demonstrate differences in assumptions while highlighting advantages and limitations of each method. Additionally, we provide guidance on selecting the most appropriate statistical method based on the scale of the data and intended inference. To illustrate the varying ecological insights derived from each statistical model, we apply them to the movement track of a single ringed seal (*Pusa hispida*) in a case study. Through our case study, we demonstrate that each model yields varying ecological insights. For example, while the selection coefficient values from RSFs appear to show a stronger positive relationship with prey diversity than those of the SSFs, when we accounted for the autocorrelation in the data none of these relationships with prey diversity were statistically significant. Furthermore, the HMM reveals variable associations with prey diversity across different behaviors, for example, a positive relationship between prey diversity and a slow-movement behaviour. Notably, the three models identified different "important" areas. This case study highlights the critical significance of selecting the appropriate model as an essential step in the process of identifying species-habitat relationships and specific areas of importance. Our comprehensive review provides the foundational information required for making informed decisions when choosing the most suitable statistical methods to address specific questions, such as identifying expansive corridors or protected zones, understanding movement patterns, or studying behaviours. In addition, this study informs researchers with the necessary tools to apply these methods effectively.

**Keywords:** animal movement, biologging, habitat selection, hidden Markov models, integrated step-selection functions, movement ecology, Poisson point process models, resource selection functions, ringed seal (*Pusa hispida*), step-selection functions, telemetry

## 1. Background

Understanding the relationships between animals' space use and their physical environment is a fundamental aspect of ecological research and is essential for species conservation [1–3]. Movement data from biologging devices provide valuable information on animal space use [4,5], and statistical models are often used to relate the movement data (in terms of occurrence, movement, or behaviour) to indicators of resources (e.g., vegetation [6]), proxies of energy (e.g., terrain ruggedness [7]), or perceived predation risk (e.g., canopy cover, [8]). These models can provide insight into the environmental conditions associated with key ecological concepts such as home range, habitat selection, movement corridors, behaviour, and critical habitat [5,9]. Identifying and designating critical habitat is one of the main ways governments protect species from rapid and widespread habitat degradation and climate changes (e.g., under the Environment Protection and Biodiversity Conservation (EPBC) Act in Australia, Species at Risk Act (SARA) in Canada, Wildlife and Countryside Act in the United Kingdom, Endangered Species Act (ESA) in the United States). Thus, it is essential that statistical models that use movement data to identify critical habitat are chosen, implemented, and interpreted properly.

Various models aim to link animal movement data to environmental covariates, but each model is appropriate for specific research questions [10,11]. Many of these models can be easily implemented using R packages (e.g., `amt`, [12]; `momentuHMM`, [13]; see [14] for a relevant review of packages), and their use is ubiquitous in animal movement literature (e.g., [15,16]). However, choosing an appropriate method is a complex and challenging aspect of movement modeling [14]. For example, different models address ecological questions at different scales, from large-scale questions on important areas for species (e.g., first- or second-order selection at the species or home range scale, respectively), to smaller-scale movement- or behaviour-specific questions (e.g., third- or fourth-order selection at the habitat usage or actual food intake scale, respectively; [17–19]). Some methods are designed to understand habitat preference via selection functions (e.g., resource selection function [RSF], step-selection function [SSF], or integrated-SSF) and others are focused on understanding how habitat relates to discrete behavioural states (e.g., hidden Markov model [HMM]). As such, consideration of scale and the behaviour of interest when choosing a model is imperative for meaningful interpretation of results.

Here, we describe and compare three mainstream, and readily implemented, models for linking animals' movement data to environmental covariates: RSF, SSF, and HMM. While these models answer different ecological questions and require different resolutions of data, they are all commonly applied to characterize relationships between species and their environments using movement data. Our objective is to help clarify the intended use and mathematical underpinnings of each model to help ecologists properly choose between them. For example, RSFs and SSFs are typically used to address similar questions on habitat selection, yet SSFs generally require relatively high-frequency data compared to RSFs. SSFs and HMMs generally both require movement data at a finer temporal resolution, yet SSFs are commonly used to address questions regarding movement and habitat selection whereas HMMs are used to link specific animal behaviours to environmental covariates. Reviews comparing selection functions (e.g., [20,21]) or selection functions with inhomogeneous point process models (IPPs, e.g., [22]) are available, but here, our review is unique as we compare selection functions and HMMs. HMMs are a fundamentally different, but increasingly popular, model to relate animal movement data to environmental covariates, and thus, a review of RSFs, SSFs, and HMMs is warranted.

**Table 1**. Terminology used in this paper.

| **Term** | **Definition** |
|---|---|
| Habitat | The set of environmental covariates (biotic resources and abiotic conditions) that characterize the space an animal inhabits [23] |
| Habitat selection | The process whereby individuals preferentially use, or occupy habitats [24] |
| Habitat selection function | A function proportional to the probability of selection of habitat [23] |
| Habitat availability | The accessibility, prevalence, and procurability of physical and biological components of a habitat by animals [25] |
| Habitat use | The exploitation of habitat to meet a biological need [26]; in the RSF analysis, the presence of an animal in a habitat [23] |

## 2. Description of statistical models

### *2.1 Resource selection functions (RSF)*

A resource selection function (RSF) is a widely used function that relates habitat characteristics to the *relative* probability of use by an animal [27–30]. RSFs can assume that the environmental conditions selected by an animal provide desirable resources, and that the probability of habitat selection (Table 1) is a function of predictor variables representing a resource distribution on a landscape [31]. Although RSFs are now often referred to as *habitat* selection functions as they can include covariates other than resources (e.g., predator probability; [32]).

When applied to movement data, RSFs often compare observed animal locations, which represent a sample of the habitat used by the animal, to a sample of locations randomly selected within an animal's home range (e.g., the minimum convex polygon (MCP), or extent, of the observed locations, Fig. 1). Observed locations are often referred to as *used* locations, while the random background locations are often referred to as *available* locations [30]. RSFs are widely applied in part due to their ease of use (e.g., using the `amt` R package, [12]) and that they can provide broad scale information on species-habitat relationships [21].

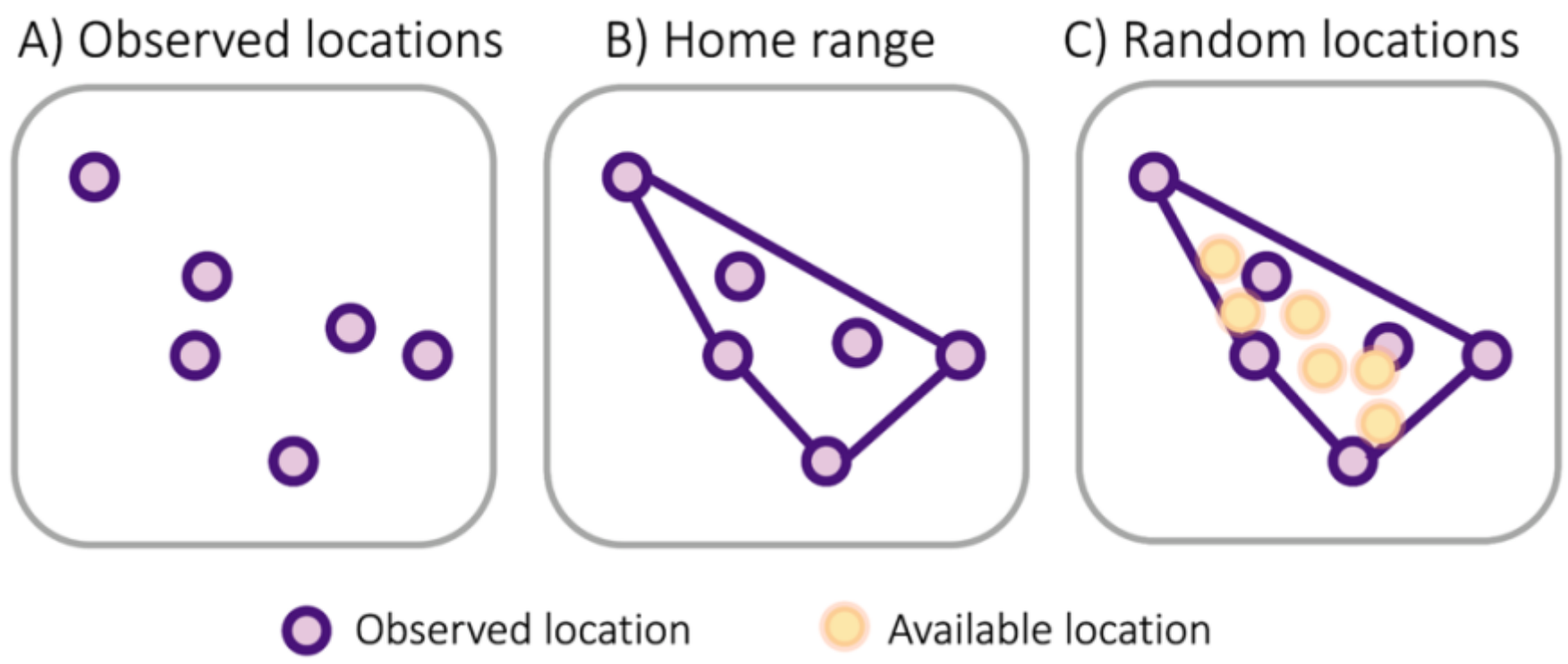

**Figure 1**. A) observed (i.e., “used”) animal locations, B) the home range (here via minimum convex polygon), and C) available locations randomly sampled from within the minimum convex polygon.

The RSF, $w(\mathbf{x})$, encodes the relative strength with which an animal selects a given set of habitat covariate values $\mathbf{x}$ over all habitat covariate values available to them (e.g., one of these covariates $\mathbf{x}$ may be the density of food at a location [23,29,38,39]). The RSF, $w(\mathbf{x})$, is typically defined and interpreted in the exponential form:

**Equation 1**

$$w(\mathbf{x}) = \exp(\beta_1 x_1 + \beta_2 x_2 + \cdots + \beta_k x_k),$$

where $\mathbf{x} = \{x_1, \ldots, x_k\}$ denotes the values of $k$ predictor habitat variables and $\beta_1, \ldots, \beta_k$ are the associated selection coefficients [27]. In practice, the coefficients of the RSF ($\beta_1, \ldots, \beta_k$) are commonly estimated using a logistic regression. Namely, suppose that a total of $n$ resource units are monitored, and the response $\mathbf{y} = \{y_1, \ldots, y_n\}$ is a set of binary random variables representing observed ($y_i = 1$) and available ($y_i = 0$) resource units. Then, the probability that resource unit $i$ is used given its environmental covariates $\mathbf{x}_i = \{x_{1,i}, \ldots, x_{k,i}\}$ is modeled as:

**Equation 2**

$$P(y_i = 1|\mathbf{x}_i) = \frac{\exp(\beta_1 x_{1,i} + \beta_2 x_{2,i} + \cdots + \beta_k x_{k,i})}{1 + \exp(\beta_1 x_{1,i} + \beta_2 x_{2,i} + \cdots + \beta_k x_{k,i})}.$$

The RSF from eqns 1-2 is written in environmental space, but RSFs can also be written in geographic space as an inhomogeneous Poisson point process (IPP). RSFs written as IPPs describe the density of points (e.g., animal locations) across all physical locations available to the animal (e.g., *s* may be the easting and northing coordinate of an animal in eqn 3) [23, 40,41]. An important difference between the two is that RSFs written in environmental space generally model directly the relationships with environmental covariates, $w(\mathbf{x})$, while IPPs account for the relationship with environmental covariates through modelling the density of points [30,42]. Specifically, the IPP models the density of points as the exponential of a linear function of spatial predictors – interpreted as the intensity function, $\lambda(s)$:

**Equation 3**

$$\lambda(s) = \exp(\beta_0 + \beta_1 x_1(s) + \beta_2 x_2(s) + \ldots + \beta_k x_k(s)),$$

where *s* is the location in geographical space, $x_1(s)$, …, $x_k(s)$ are *k* predictor habitat variables associated with location *s*, $\beta_0$ is an intercept term, and $\beta_1$, …, $\beta_k$ , are the selection coefficients similar to eqn 1. The intensity function $\lambda(s)$ is proportional to $w(\mathbf{x})$ from eqn 1 and the selection coefficients can be estimated with a logistic regression. As the number of available points sampled for the logistic regression grows to infinity, the selection coefficients ($\beta_1$, …, $\beta_k$) from eqn 2 (which are estimated using logistic regression) converge to the selection coefficients ($\beta_1$, …, $\beta_k$) from eqn 3 and the IPP model [42]. Note that if the model intercept $\beta_0$ is estimated it must be interpreted with care as it is often not biologically meaningful (i.e., it is related to the density of observations, but for movement data that may reflect frequency of observations; [30,39]).

While the RSF $w(\mathbf{x})$ represents selection, it more formally represents the ratio between the use (i.e., utilization) distribution $F^U(\mathbf{x})$ (the frequency distribution of habitat covariates used by an animal) and available distribution $F^A(\mathbf{x})$ (the frequency distribution of habitat covariates assumed to be available) (i.e., $w(\mathbf{x}) \propto F^U(\mathbf{x})/F^A(\mathbf{x})$). This relationship can be seen from the following definition of the use distribution, $F^U(\mathbf{x})$:

**Equation 4**

$$F^U(\mathbf{x}) = \frac{w(\mathbf{x})F^A(\mathbf{x})}{\int_{\mathbf{x}' \epsilon \Omega} w(\mathbf{x}')F^A(\mathbf{x}')d\mathbf{x}'},$$

where in the denominator we integrate over the entire domain of the environment, $\Omega$, to ensure that $F^U(\mathbf{x})$ is a valid probability distribution that integrates to one. Equation 4 shows that the use distribution is a function of both the available habitat (as defined by $F^A(\mathbf{x})$) and the selection function, $w(\mathbf{x})$[23,29,38,39]. As movement datasets only contain information on used habitat variables, available resource units, which should not erroneously be treated as true absences, must be sampled from $F^A$. A similar definition of the use distribution can be made in geographical space:

**Equation 5**

$$F^U(s) = \frac{w(\mathbf{x}(s))\, F^A(s)}{\int_{s' \epsilon\, G} w(\mathbf{x}(s'))F^A(s')ds'},$$

where the selection function $w(\mathbf{x}(s))$ is now a function of $\mathbf{x}(s)$, the habitat covariates $\mathbf{x}$ at location $s$, and the available distribution $F^A(s)$ is a function of location and is generally assumed to be constant (i.e., $F^A(s) = \frac{1}{|G|}$; [30]). In the denominator, we integrate over the geographical area available to the animal $G$ [43].

RSFs' assumptions of independence and large-scale availability make them particularly well-suited for analyzing coarse animal movement data to understand the broad-scale spatial ecology of animals. For example, RSFs have been used to identify broad conservation corridors [33] and high wildlife density areas to improve rabies vaccination programs [34]. Additionally, RSFs have been used to understand how habitat selection is affected by seasons [35] anthropogenic developments [36], and the presence of predators [32], and can be used to understand the cumulative effects of such factors [37].

When applied to movement data, simple RSFs are most suitable for data collected at a coarse temporal resolution. Autocorrelation in movement data can introduce spatio-temporal correlation that cannot be explained by habitat-related covariates. If not accounted for, this autocorrelation can result in underestimation of standard errors, potentially inflating Type I error rates [23,51]. This issue is particularly common when using high-resolution datasets [52] and is expected to become more significant with the advancement of technology that enables increasingly fine-resolution movement data [53]. Various approaches have been proposed to mitigate this problem, including data thinning [52], weighting points [54], correcting the standard errors [55], and adding an autocorrelation term (e.g., intrinsic conditional autoregressive (ICAR) model, [41]).

### *2.2 Step selection functions (SSF)*

A step selection function (SSF) builds on RSFs and estimates an animals' resource selection at each observed sequential location or *step* (i.e., the linear segment between two consecutive locations; Fig. 2) based on both habitat covariates and movement constraints [44,58,59], partially providing a solution to the autocorrelation issues associated with RSFs (but see [60,61]). Unlike RSFs, which assume the entire home range is available over the study period, SSFs assume that an animal can only access a limited area near its previous location within the time scale of a step. There are many versions of SSFs, including many versions that focus simply on the influence of environmental covariates on selection. In our review, we focus on a commonly applied version often referred to as integrated-SSFs [56] because it allows for environmental covariates to influence animal movement itself in addition to selection. These SSFs allow the movement characteristics (including step length and turning angle) and environmental covariates to be combined in a single linear predictor, which allows for joint estimation, and potential interactions, between movement and environment, thereby relaxing the assumption that observed movement attributes (i.e., velocities and directional persistence) are independent of resource selection [56,57]. That is, these SSFs simultaneously estimate the habitat selection and movement kernel. The movement kernel describes how an animal would move in the absence of habitat selection and thus what is available to the animal. In these SSFs, where environmental covariates influence both animal movement and selection, availability is estimated simultaneously with habitat selection through the movement kernel, rather than being assumed a priori [45,56,59].

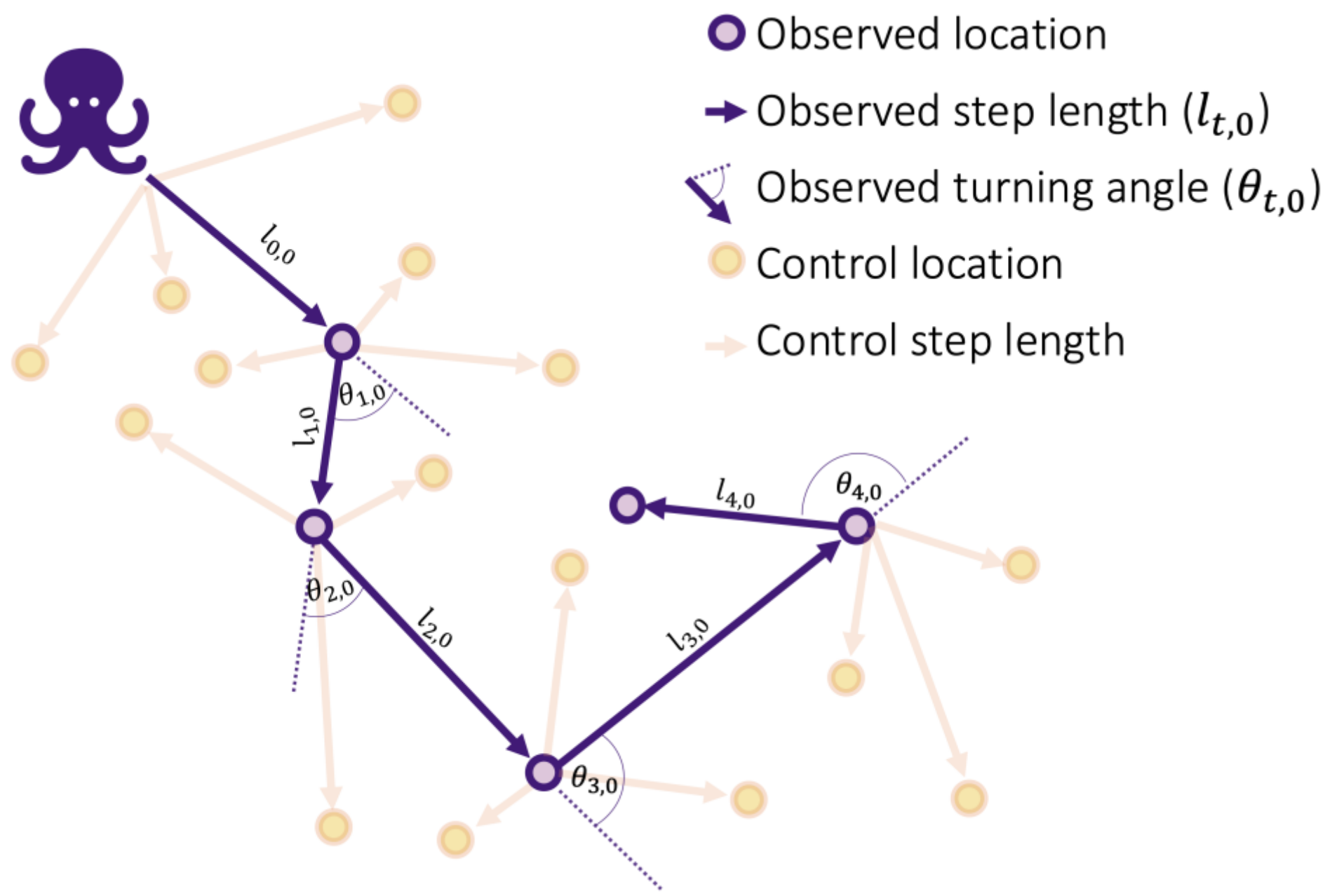

**Figure 2.** Example of an animal movement track with linear *observed steps* between locations. An observed turning angle is relative to the trajectory from the preceding step. At each observed location, random *control* locations (m = 3; and their associated step lengths and turning angles) are generated to estimate availability.

Step selection functions model habitat selection and movement explicitly, by defining use distribution as the product of the habitat selection $w$ and a selection-free movement kernel $\phi$ [56,57]. The movement kernel is essentially equivalent to having the availability distribution of an RSF, $F^A(\mathbf{x})$, vary temporally, and it represents how the animal would move in the absence of habitat selection [56,57]. Thus, the probability density of an animal using location $s_t$ at time $t$ given its previous locations $\mathbf{s}_{1:t-1} = \{s_1, s_2, \dots, s_{t-1}\}$ is modeled as:

**Equation 6**

$$F^U(s_t|\mathbf{s}_{1:t-1}) = \frac{w(s_{t-1}, s_t)\phi(s_t, \mathbf{s}_{1:t-1})}{\int_{s' \in G} w(s_{t-1}, s')\phi(s', \mathbf{s}_{1:t-1})ds'},$$

where $w(s_{t-1}, s_t)$ is the habitat selection function and $\phi(s_t\ , \mathbf{s}_{1:t-1})$ is the selection-free movement kernel (although this terminology is inconsistently used, see [45]). As in eqn 5, the denominator integrates over the entire geographic domain $G$ to normalize the use distribution. Step selection functions typically require observed locations to be at a constant time interval, and the parameters of the SSF and movement kernel are dependent on time index $t$ [56,57]. Similar to RSF coefficient estimation with a logistic regression, SSFs coefficients are often estimated by fitting a conditional logistic regression. SSFs coefficients can also be estimated using Monte Carlo integration techniques [45].

Fitting a conditional logistic regression to estimate the SSF coefficients generally entails drawing $m$ *control* points at each time *t*. We denote the vector with the locations of observed and control points as $\mathbf{s}_{t,0:m} = \{s_{t,0}, s_{t,1}, \dots, s_{t,m}\}$, where $s_{t,0}$ is the location (i.e., the coordinates) of the observed point at time *t* and $s_{t,i}$ for $i = 1, \dots, m$ is the location of the $i^{\text{th}}$ control point. Further, $y_{t,i}$ is a binary variable that equals 1 when $i = 0$ (observed) and 0 when $i = 1, \dots, m$ (control). Control points (also known as random, sample, or integration points) are often generated using observed step length and turning angle distributions (Fig. 2; [45]). While such control points can be misinterpreted as solely representing available movement options, they are often used as a computationally efficient tool to estimate availability at high resolution close to the animal's last location [45]. The conditional probability that the point at location $s_{t,i}$ is the observed location given the environmental covariates at that location $\mathbf{x}_{t,i}$ and the animal's previous locations $\mathbf{s}_{1:t-1,0}$ [45,56] is:

**Equation 7**

$$P\left(y_{t,i} = 1 \middle| \mathbf{X}_t, \mathbf{s}_{t,0:m}, \mathbf{s}_{1:t-1,0}\right) = \frac{\exp(\eta_{t,i})}{\sum_{j=0}^{m} \exp(\eta_{t,j})},$$

where $\mathbf{x}_{t,i}$ is row $i$ of the matrix $\mathbf{X}_t$ and $\eta_{t,i}$ is a linear predictor that arises from the fact that both the habitat selection function and the movement kernel of eqn 6 can take an exponential form. The linear predictor typically includes a vector of $k$ environmental covariate values $\mathbf{x}_{t,i}$ at location $s_{t,i}$ and movement variables defined by previous and current locations, such as the natural log of step length and cosine of turning angle as covariates (e.g., to model step length with a gamma distribution and turning angle with a von Mises distribution [45,56]). To model a movement kernel that can change as a function of environmental covariates (e.g., slower walking speed in deep snow), $\eta_{t,i}$ can include interactions between movement variables and the environmental covariates, including, for example, environmental covariates at the previous observed location $s_{t-1,0}$ . For example, the linear predictor can be expressed as:

**Equation 8**

$$\eta_{t,i} = \beta_1 x_{1,t,i} + \beta_l l_{t,i} + \beta_{\ln} \ln(l_{t,i}) + \beta_\theta \cos\left(\theta_{t,i}\right) + \beta_2\, x_{2,t-1,0} \ln\left(l_{t,i}\right),$$

where $x_{1,t,i}$ represents an environmental covariate at location $s_{t,i}$ that affects selection (e.g., food density), $x_{2,t-1,0}$ an environmental covariate at location $s_{t-1,0}$ that affects movement itself (e.g., snow depth), $l_{t,i}$ the step length between observed location $s_{t-1,0}$ and potential location $s_{t,i}$, $\theta_{t,i}$ the turning angle between observed locations $\mathbf{s}_{t-2:t-1,0}$ and potential location $s_{t,i}$, and the $\beta$s are coefficients to estimate. This formulation allows SSFs to integrate both movement dynamics and environmental influences into step-selection probability estimates.

Step selection functions are well-suited for fine-scale movement datasets and have been used for understanding how environmental covariates affect animal movement characteristics [62]. For example, SSFs have been used to show how roads increase step lengths [63]. SSFs have also been used to understand how predators or social networks influence movement [58,64]. Further, SSFs have been used to determine how variation in temperature affects movement and use of thermal cover [15]. Additionally, SSFs have been used to understand how slopes associated with thermal uplift affect soaring bird movement patterns, including potential stopover movements to access carrion [65].

### *2.3 Hidden Markov models (HMM)*

In the context of movement ecology, hidden Markov models (HMMs) are principally used to classify movement into a finite number of discrete behavioural states (e.g., searching, resting, traveling, [66]), and differ conceptually from RSFs or SSFs as they generally do not make inference on habitat selection. However, HMMs can be used to investigate relationships between environmental covariates and animal movement and behaviour [67,68].

HMMs assume there is a hidden state process unfolding over time from which we obtain observation data (Fig. 3). That is, it is assumed that at any time, animals exhibit one of $N$ discrete and temporally autocorrelated states $Z_t \in \{1, 2, \ldots, N\}$, which represent hidden (unobserved) behaviours. These states are related to an observation time series ($Y_1, \ldots, Y_T$), where $T$ is the total number of time steps. In particular, given the hidden state process, $Z_1, \ldots, Z_T$, each observation, $Y_t$, is assumed to depend only on its corresponding hidden state, $Z_t$ (Fig. 3). In contrast to the binary (observed/available) values used in the observation time series of RSFs and SSFs, the observation used in animal movement HMMs, generally comprise time series of step lengths and turning angles [73,74] and can include additional data streams such as dive depth or acceleration [75].

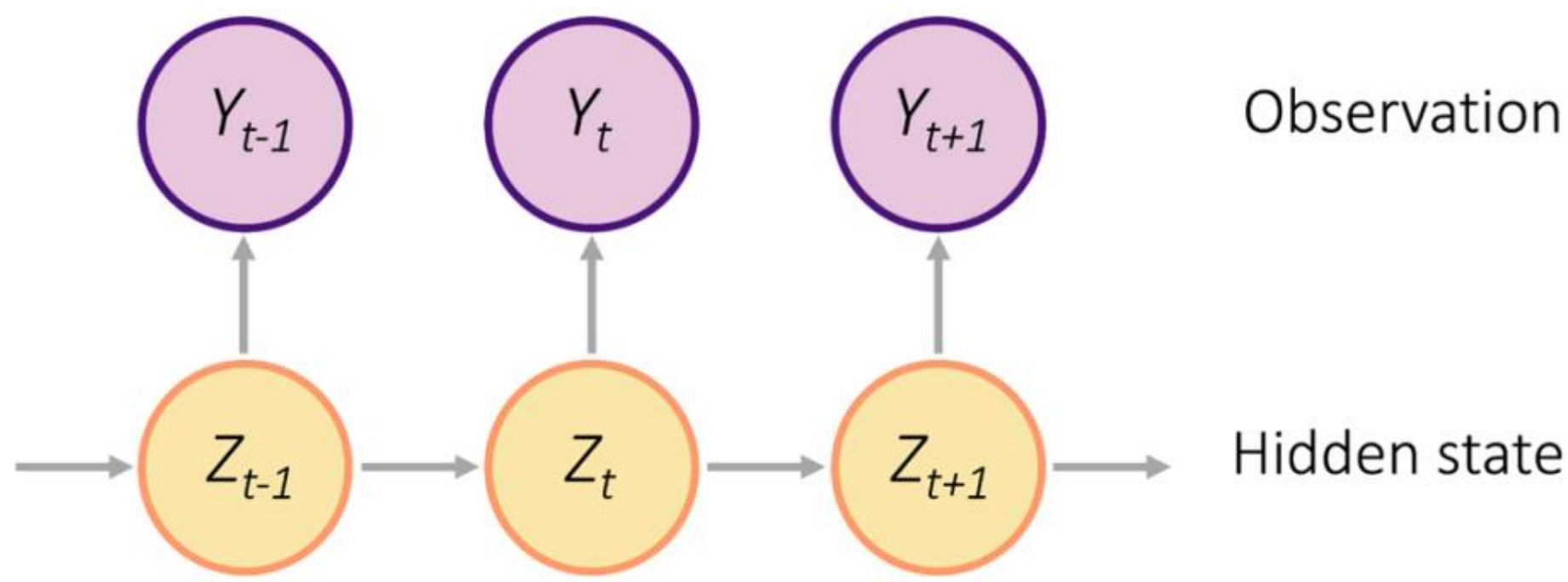

**Figure 3**. The structure of a hidden Markov model, where the hidden states ($Z_t$) depend on the previous state, and the observed data ($Y_t$) depends on the hidden state. In the case of modeling animal movement, the observed process is composed of empirical data such as step lengths and turning angles, and the hidden state may represent behaviours such as foraging, resting, and traveling.

For an *N*-state HMM (e.g., *N* = 3 states, representing foraging, resting, and traveling), the state transition probability matrix (**Γ**) is an *N* x *N* matrix, where entry (*i*,*j*) is denoted as $\gamma_{i,j}$ and represents the probability of transitioning from state *i* to state *j*. Namely, we have $\gamma_{i,j} = Pr(Z_{t+1} = j \mid Z_t = i)$, and each row of **Γ** must sum to 1 (i.e., $\sum_{j=1}^{N} \gamma_{i,j} = 1$). All observations are assumed to be independent of one another after conditioning on the underlying hidden state process: $Pr(y_t | Z_1, \ldots, Z_t;\, y_1, \ldots, y_{t-1}, y_{t+1}, \ldots, y_T) = Pr\,(y_t | \, Z_t)$ . We define an *N* x *N* diagonal matrix, $\mathbf{P}(y_t) = diag(p_1(y_t),\ldots,p_N(y_t))$, where $p_i(y_t)$ represents the probability density of observation $y_t$ given state $i$ (i.e., $p_i(y_t) = Pr(Y_t = y_t | Z_t = i)$). When modeling animal movement, step length is often modeled with distributions such as gamma, exponential, or Weibull, while turning angle is modeled with circular distribution such as the von Mises or wrapped Cauchy [74].

Taken together, the HMM likelihood function ($L$) is:

**Equation 9**

$$L = \boldsymbol{\delta}\mathbf{P}(y_1)\boldsymbol{\Gamma}\mathbf{P}(y_2)\boldsymbol{\Gamma}\mathbf{P}(y_3),\ldots,\boldsymbol{\Gamma}\mathbf{P}(y_T)\mathbf{1}',$$

where $\boldsymbol{\delta}$ is the initial state distribution of the Markov chain, and $\mathbf{1}'$ is a row vector of ones. The parameters can be estimated using numerical methods to maximize the likelihood (eqn 9) or with the expectation-maximization (EM) algorithm [68].

Environmental covariates can be incorporated into two different parts of an HMM. First, to examine how covariates affect the probability to transition between states (e.g., how vegetation density may increase the probability to transition into foraging for an herbivore), they can be used to alter the transition probabilities, $\gamma_{i,j}$. Specifically, we can make transition probabilities, now denoted as $\gamma_{t,i,j}$, change at each time *t* as a function of a vector of environmental covariates $\mathbf{x}_t$, e.g., via a multinomial logit link:

**Equation 10**

$$\gamma_{t,i,j} = \frac{\exp(\eta_{t,i,j})}{\sum_{l=1}^{N}\exp(\eta_{t,i,l})},$$

and

**Equation 11**

$$\eta_{t,i,j} = \beta_{0,i,j} + \beta_{1,i,j} x_{1,t} + \ldots + \beta_{k,i,j} x_{k,t} \ ,$$

where $\beta_{0,i,j}$ is an intercept term and $\beta_{1,i,j}, \ldots, \beta_{k,i,j}$ are the behaviour-specific regression coefficients for each $k$ environmental covariates with values $\mathbf{x}_t = x_{1,t}, \ldots, x_{k,t}$ at time $t$ [69,73,76]. To ensure model identifiability, $\beta_{0,i,j}, \beta_{1,i,j}, \ldots, \beta_{k,i,j}$ are fixed to 0 for one element in each row of the state transition probability matrix, typically, the diagonal when $i = j$.

Second, environmental covariates (typically abiotic) can be used to alter the observation probability density function, $p_i(x)$, which can be interpreted as examining how they affect the movement of the animal within state $i$ (e.g., how snow depth affects the speed of a walking animal). To do so, the observation probability density functions will change through time, now denoted $p_{t,i}(x)$, and their parameters will depend on environmental covariates. For example, environmental covariates can affect step length mean $\mu_{t,i}^{(l)}$ for each state *i* at each time step *t* as

**Equation 12**

$$ln(\mu_{t,i}^{(l)}) = \beta_{0,i}^{(\mu^{(l)})} + \beta_{1,i}^{(\mu^{(l)})} x_{1,t} + \ldots + \beta_{k,i}^{(\mu^{(l)})} x_{k,t} \ .$$

Similarly, environmental covariates can affect other observation parameters such as the standard deviation of the step length distribution and turning angle concentration (see Appendix 2.3). Additionally, movement can be modelled with an orientation bias (i.e., biased random walks) or further, with bias toward any angle relative to stimulus (e.g., olfactory search perpendicular to wind; [71]).

Hidden Markov models are specific types of state-space models [77], where the states are discrete (i.e., there is a finite number of hidden states). Other state-space models have been developed that describe animal movement behaviour as a continuum of move-persistence ranging from 0 to 1. In these models, the state represents the move persistence where values closer to 0 are indicative of less move persistence (i.e., area-restricted search), and values closer to 1 are indicative of more persistent movement (i.e., traveling; [78]). However, these state-space models assume that behaviour exists on a linear continuum of only two behaviours (e.g., area-restricted search to traveling; e.g., [79]), and it is unclear where other behaviours (e.g., resting) would fall along that continuum [80]. Thus, state-space models with discrete state space such as HMMs may be easier to interpret, as each state has defined respective turning angle and step length distributions [80].

Hidden Markov models are well-suited for high-frequency movement datasets and behaviour-specific research questions since they provide an approach to identify habitat features relevant to separate behaviours [69]. For example, HMMs have been used to quantify how foraging probability relates to canopy cover [70], how travel speed relates to snow depth [16], and how animal orientation relates to wind direction [71]. Further, HMMs have been used to determine how variation in human influence on the landscape affects movement mode [72].

## 3. Choosing a model

By design, RSFs are well-suited to address large-scale questions on important areas for species (e.g., second-order selection; at the home range scale), whereas SSFs and HMMs are typically well-suited to address smaller-scale movement- or behaviour-specific questions (Box 1, e.g., third-order selection; at the habitat usage scale; [17–19,81]). Since RSFs often generate the availability sample within the home range (MCP) of the observed data, and SSFs utilize a movement kernel that constrains the model to the scale of a step [56], the SSF prediction coefficients provide insight at a smaller scale than in the large-scale (home range) case of RSFs (although see [82–84] for guidance on scaling up from SSFs). In contrast, HMMs classify movement into discrete behavioural states, and thus do not consider habitat covariates as a driver of habitat selection, rather, as a mechanism related to estimated behavioural states [67,68].

**BOX 1**. **Which model to choose?**

Decision tree for choosing an appropriate modeling approach for understanding animal relationships with environmental conditions using movement data, including if you are interested in how the environmental conditions influence animal space use or movement, and if the data are coarse- or fine-scale. For example, caribou movement in the winter could be related to snow depth. RSFs can be used for broad-scale inference relating caribou occupancy and snow depth, SSFs can be used to understand whether snow depth affects selection at finer scale, including if it slows down movement speed, and HMMs can identify the depths of snow associated with a behaviour (e.g., foraging).

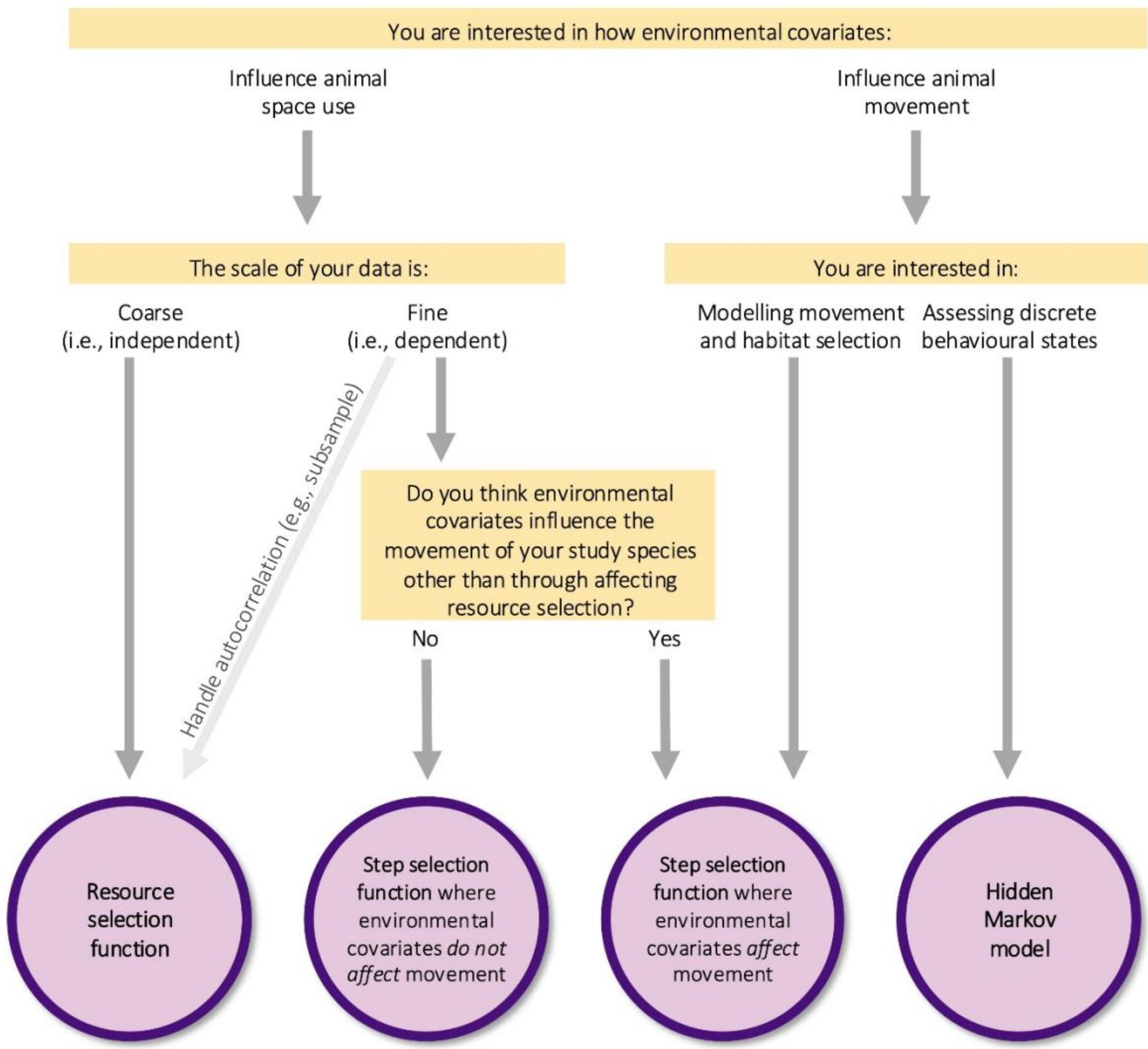

### *3.1. Important considerations*

Choosing an appropriate model can be difficult, and choosing an adequate model does not guarantee appropriate inference. For example, [86] used HMMs to identify white tailed deer (*Odocoileus virginianus*) male-female interaction events, but upon validation, found that the model incorrectly identified interactions, and its use would mislead researchers about when and where mating behaviours are occurring. Additionally, [87] used move-persistence models to identify relationships between foraging behaviour and prey abundance, and found mis-matches between foraging behaviour, prey biomass, and diving behaviour. These cautionary tales highlight the importance of understanding the chosen models' design, abilities, and limitations (Box 2). Where possible, validation exercises (e.g., leave-one-individual-out cross-validation, [88]) and incorporation of additional data streams (e.g., dive data, [79]) is recommended.

Selection functions are subject to several assumptions. Selection functions assume that animals make resource-driven choices where at each location they choose the best habitat to move to next, but this may not always be the case (e.g., as in migration, or when relying on memory but see [89]). RSFs assume that locations are independent, and methods have been proposed to circumvent the violation of this assumption (e.g., data thinning [21], weighted approach [54]). Similarly, SSFs assume that the step length and turning angle are independent, yet autocorrelation may exist depending on the scale (e.g., animals may keep foraging in nearby locations; [73,90]; but see examples to account for spatial and temporal structure: [91–93]). Thus, sampling intervals need to be examined because small time intervals can lead to misleadingly narrow confidence intervals [54,56]. Additionally, SSFs assume independence between step length and turning angle, yet cross-correlation between the two may be unavoidable in some animal movement datasets [73]. Solutions to address this issue have been proposed [94,95]. While many RSF and SSF models assume that individuals exhibit identical movement and selection, individual variation can be included in selection functions via random slopes in a mixed model (e.g., using the `glmmTMB` or `mgcv` package in R, [48,96]) or by using a two-step approach of fitting individual models and averaging parameter estimates for inference at the population level (e.g., using the `TwoStepClogit` package in R, [53,62]).

Choosing the number and the spatial extent of the availability sample are important computational and ecological considerations, respectively, and can affect the coefficient estimates and subsequently the ecological interpretation [23,44,45]. The number of available samples can be based on a constant (e.g., 10,000, [31,38,46]), point density (e.g., 1 available location per $km^2$, [47]), or by a ratio of observed:available locations (e.g., 1:20, [23]). Nevertheless, it is advisable to apply the selection function using a range of available sample sizes and choosing the sample size where the coefficient estimates stabilize [23,30,42,46]. As approximation accuracy increases with the size of the available sample, it is generally better to include a larger availability sample, especially given quick computation time [23,45,48]. For RSFs, the spatial extent of the availability sample represents the home range of the animal, which is often represented as the MCP (e.g., absolute, buffered, or clipped) of the observed sample and can affect the scale of inference and thus should be chosen after consideration of the species' ecology (see [49,50]).

Movement HMMs rely on different assumptions that relax many assumptions of selection functions (RSF, SSF). Unlike selection functions, movement HMMs consider habitat covariates as related to estimated behavioural states, rather than as a driver of habitat selection. Nonetheless, HMMs acknowledge that telemetry data encompass distinct, unobserved behaviours and distinguish the state process from the observation process. HMMs also acknowledge that telemetry data are not independent, and explicitly model autocorrelation through the state process. Movement patterns and habitat associations are behaviour-specific, which can be modeled by the state-dependent distribution of the observation process and state transition probabilities.

These models (RSFs, SSFs, and HMMs) can be easily extended to incorporate inter- or intra-specific relationships. For example, interspecific relationships can be included by adding another species' probability of occupancy as a covariate [32,48,97], or by using an individual-based model [98]. Additionally, more complex and dynamic intraspecific relationships can be included by considering individual animal's movement relative to the group movement (see [99] for an HMM example).

Maps of probability of use are most easily generated using RSFs [28] and are particularly useful for conservation management and land-use planning [33,100,101]. Maps can be made using SSFs and HMMs [12,13,82–84], however for SSFs, require path simulation exercises with setting choices that are subjective (e.g., choosing the number of tracks and steps in the simulation, the burn-in size). Recent work has offered guidance on space use predictions from SSFs, including implementation in `amt` [12,83,84] (and see our step-by-step tutorial in Appendix 4).

**BOX 2**. **Model comparison chart.** A summary comparison of statistical models that use movement data to characterize species-habitat associations. Note that this table highlights the main intended use, assumptions, strengths, and weaknesses associated with each model and their implementation and is not an exhaustive list.

| **Intended use(s)** | **Main assumption(s)** | **Strengths** | **Considerations in implementation** |
|---|---|---|---|
| **Resource selection function (RSF)** | | | |
| Relate environmental covariates to animal occurrence, typically at home-range (large) scale | Habitat selection depends on the encountered environmental conditions;<br><br>The probability of selection is constant during the period of investigation[1];<br><br>Locations are independent[2] | Simple to use and interpret;<br><br>Popular (can compare results to other studies) | Challenges in choosing the quantity of points for the availability sample and the extents of the available sample[3];<br><br>Need to potentially consider potential movement barriers (i.e., is the complete home range is assumed to be available to the animal) |
| **Step selection function (SSF)** | | | |
| Relate environmental covariates to animal occurrence at a movement (small) scale; | Habitat selection is conditional on occurrence within a restricted range (i.e., available step-length is limited by observed movement); | Relatively easy to use;<br><br>Partially accounts for autocorrelation;<br><br>Creates availability samples based on | Challenges in choosing the quantity of points for the availability sample[3];<br><br>Selection of interval between movement locations must meet assumptions of independence; |
| Can assess how movement characteristics (i.e., step length) are affected by environmental covariates[↟] | Movement can be affected by environmental conditions[↟]<br><br>Steps are independent | observed movement patterns;<br><br>Allows for simultaneous inference on both movement and habitat selection[↟] | Scale of steps dictates habitat availability and the types of behaviours that can be modeled;<br><br>Interpretation and generation of prediction maps is more nuanced because selection is relative to observed points (especially challenging when modeling how habitat affects movement[↟]); |
| **Hidden Markov model (HMM)** | | | |
| Assess how movement characteristics are affected by environmental covariates;<br><br>Identify environmental covariates that | Behaviour at time *t* depends only on behaviour at *t-1* (Markov property)[4];<br><br>The distribution of observations (e.g., step length and turning angle) depend on the latent | Ability to identify behavioural states and how their occurrence is associated with environmental covariates;<br><br>Ability to quantify how movement | Sampling interval dictates the types of behaviours that can be modeled;<br><br>Limited to relationship with environmental condition experienced by the animals tagged, is not designed for habitat selection |
| affect behavioural state transitions | behaviour (state) but not the previous observations[4] | characteristics (e.g. step length) are affected by environmental covariates;<br><br>State sequence can identify when and where behaviours occur | |

[1]In typical usage, but see [102].

[2]In typical usage. Solutions exist in including an autocorrelation term in the model (e.g., when using integrated nested Laplace approximation [INLA], [41]).

[3]See [23,54,102] for guidance.

[4]In typical usage, but all these assumptions have been relaxed in some form in the literature (e.g., semi-Markov model, [74], conditionally autoregressive HMM, [104], hierarchical HMM, [81]).

[†]Relevant if the SSF models the influence of habitat covariates on the movement kernel.

## 4. Case study: Ringed seals in Hudson Bay, Canada

### *4.1 Background and methods*

We used movement data of one ringed seal (*Pusa hispida*) to conduct a comparative analysis of different methods to characterize their relationship with one covariate. We chose prey diversity as the covariate of interest for exploration, given that ringed seals are known as opportunistic predators (Appendix 2; data from [105,106]). Providing an accurate description of how ringed seals interact with their environment is important because of their ecological and cultural significance. Ringed seals are facing demographic decline in some areas of their range [107] underscoring the urgency to identify important areas of occupancy for this species. In this case study, we apply each of the models to the same dataset to demonstrate and compare the insight gleaned from each. The interpretation of this case study is limited, and focuses on illustrating how the inference will be affected by which model is applied.

We analyzed the estimated movement of a seal (Fig 4a) equipped with an Argos satellite telemetry transmitter from 29 Oct 2012 to 17 Mar 2013 in Hudson Bay, Canada (see [87] for details). We used a correlated random walk state-space model (`aniMotum` R package, [108]) to filter and regularize the location data at a 24-hr time step since Argos location data is observed irregularly in time and is prone to error [109]. This procedure resulted in one location per day from 29 Oct 2012 to 17 Mar 2013 (total of 140 observed locations). Then, we matched each seal location with prey diversity in the corresponding grid cell (Fig. 4D).

We used the prey diversity data as a covariate in two RSFs (one on the full dataset and one on a thinned dataset), two SSFs (one with and one without movement-habitat interactions), and an HMM (in the transition probabilities, but see Appendix 4 for how to apply covariates to the observation probabilities). For the RSF, we included ten available locations per observed location that were sampled from within the MCP of the observed locations. We used the `amt` package in R [12] to fit the RSFs and SSFs. Specifically, we estimated the RSF coefficients associated with the prey diversity (*prey*) covariate by fitting a logistic regression to the use-available observations $y = (y_1, \ldots, y_n)$, where $y_i \in \{0,1\}$. We used a logit link function to keep the probability $P(y_i = 1|x_{preydiv,i})$ between 0 and 1:

**Equation 13**

$$P\left(y_i = 1 \middle| x_{prey,i}\right) = \frac{\exp(\beta_0 + \beta_{prey} x_{prey,i})}{1 + \exp(\beta_0 + \beta_{prey} x_{prey,i})}.$$

By default, the `fit_rsf()` function in the `amt` package calls `glm()` and thus fits logistic models with an intercept, however, in the context of an RSF based on use-available data, this intercept is not interpretable and is disregarded. Therefore, the RSF derived from this model excludes the intercept:

**Equation 14**

$$w(x_{prey,i}) = \beta_{prey} x_{prey,i} \ .$$

Since the full dataset was autocorrelated (see tutorial) and independent data is required for RSFs (see Box 1), we also fit an RSF on a thinned dataset comprising of every 10th location in the full dataset, using the same equations as the RSF on the full dataset.

We estimated the SSF coefficients by fitting a conditional logistic regression. At each time *t*, we sampled ten control locations $s_{t,i}$ for $i \in \{1, \ldots, 10\}$ from a movement kernel $\phi$ fitted to the observed data and appended them to the observed location $s_{t,0}$. Then, the conditional probability that the point at location $s_{t,i}$ is the observed location given the environmental covariates at that location $\mathbf{x}_{t,i}$ and the animal's previous locations $\mathbf{s}_{1:t-1,0}$ [45,56] is:

Equation 15

$$P\left(y_{t,i} = 1 \middle| \mathbf{X}_t, \mathbf{s}_{t,0:m}, \mathbf{s}_{1:t-1,0}\right) = \frac{\exp(\eta_{t,i})}{\sum_{j=0}^{10} \exp(\eta_{t,j})},$$

where $y_{t,i}$ is a binary variable that takes the value of 1 when the $i^{\text{th}}$ point is the observed location and 0 for the control locations and the linear predictor $\eta_{t,i}$ includes prey diversity, step length, turning angle as covariates (see Appendix 2):

Equation 16

$$\eta_{t,i} = \beta_1 x_{prey,t,i} + \beta_l l_{t,i} + \beta_{\ln} \ln(l_{t,i}) + \beta_\theta cos(\theta_{t,i}).$$

We also fit a SSF where we allowed prey diversity to affect the movement kernel (specifically, the shape of the gamma distribution), which was achieved by including an interaction between prey diversity and the natural log of the step length:

Equation 17

$$\eta_{t,i} = \beta_1 x_{prey,t,i} + \beta_l l_{t,i} + \beta_{\ln} \ln(l_{t,i}) + \beta_\theta cos(\theta_{t,i}) + \beta_2 x_{prey,t,i} \ln(l_{t,i}).$$

Note that the `fit_ssf()` function in the `amt` package calls `survival::clogit()` and thus fits conditional logistic models without an intercept. Further, we chose to include the natural log (to model the gamma distribution) of step length and the cosine (to model the von Mises distribution) of turning angle as covariates due to their conventional use in the literature and model guides [12,56,110].

Finally, we fit a three-state HMM using the `momentuHMM` package in R [13], with prey diversity as a covariate on the transition probability:

**Equation 18**

$$\gamma_{t,i,j} = \frac{\exp(\beta_{0,i,j}+\beta_{1,i,j}x_{prey,t})}{\sum_{l=1}^{N}\exp(\beta_{0,i,l}+\beta_{1,i,l}x_{prey,t})} ,$$

where $N$ is the number of behavioural states (set to $N = 3$), and $\beta_{0,i,j}$ and $\beta_{1,i,j}$ are the intercept and coefficient quantifying the effect of prey diversity on transition probability from state $i$ to $j$, respectively. We fixed $\beta_{0,i,j} = \beta_{1,i,j} = 0$ when $i = j$ (i.e., the diagonal of the transition probability matrix $\mathbf{\Gamma}$).

We predicted the probability of use relative to prey diversity (log of the relative selection strength [log-RSS] for selection functions; see [111]) and in space (steady-state utilization distributions [SSUDs]; [82]). Detailed methods are available in Appendix 2 and the data and a coding tutorial of the analysis are available in Appendix 3 and 4, respectively.

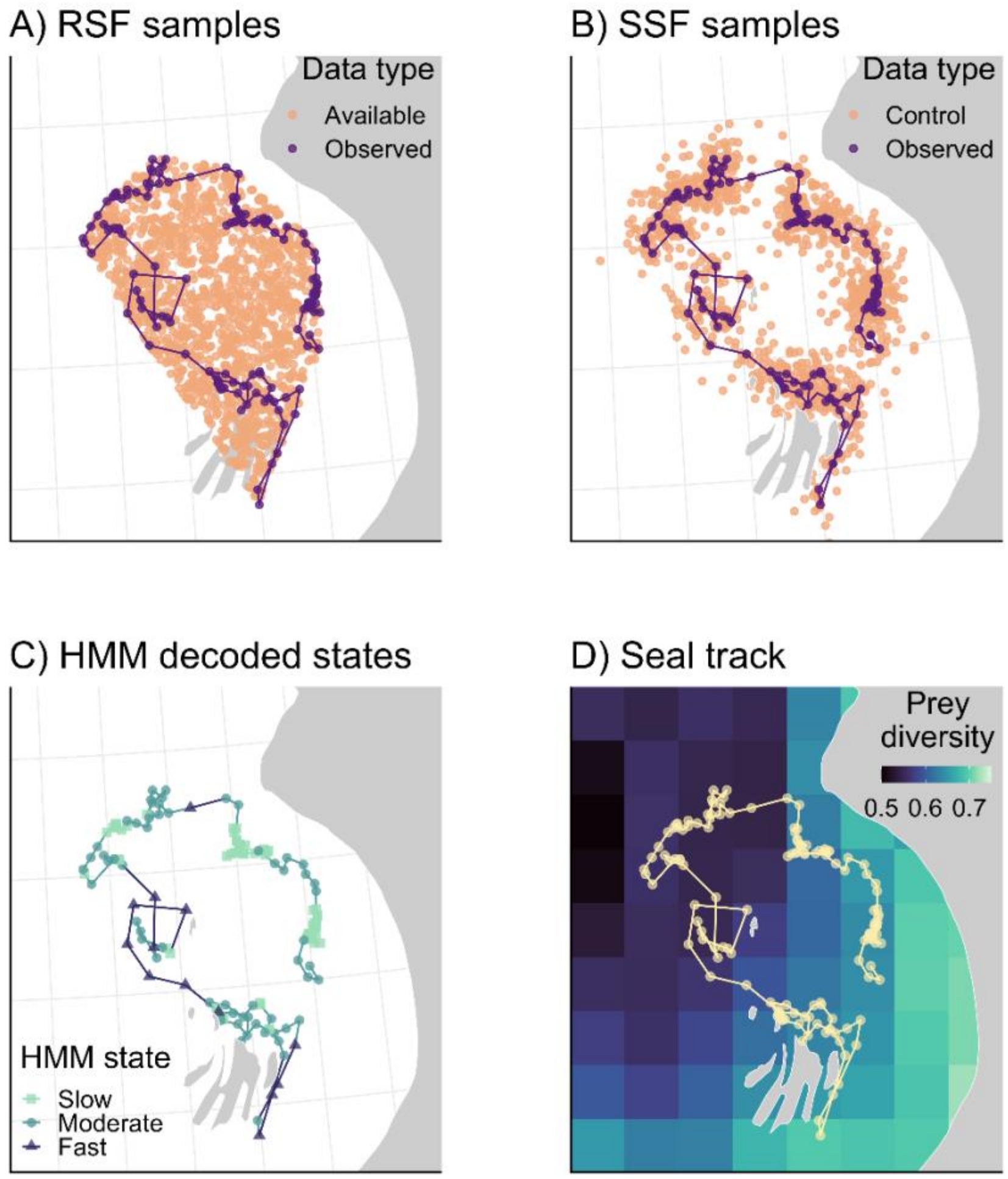

**Figure 4**. The observed (used) and available locations for the A) resource selection function (RSF) and B) step-selection functions (SSF). Purple circles represent the observed locations and purple lines denote the path between observed points. Peach circles represent the available locations for the RSF and control locations for the SSF (see Fig. 2). The RSF availability sample was generated randomly within the minimum convex polygon of the track. C) Decoded states from the hidden Markov model (HMM), including slow movement (green squares), moderate movement (blue circles), and fast movement (purple triangles). D) Ringed seal estimated locations (yellow circles) and straight-line path between locations (yellow line) overlaid on estimated prey diversity.

### *4.2 Results and discussion*

These models led to important differences in the magnitude, and sometimes direction, of the relationships with the same covariate, and thus resulted in different ecological inference. The estimated selection coefficient that represented the habitat selection relationship with prey diversity was significantly positive for the RSF on the full dataset, but not significantly different from zero for the RSF on the thinned dataset or for the SSFs (Table S1, Fig. 5A-B). Notably, the estimated coefficient for prey diversity from RSF on the thinned dataset, while not significant, was similar to that from the RSF on the full dataset, and both were much higher than that from the SSF (Table S1). The SSF with movement-habitat interactions showed no significant relationship between prey diversity and movement speed (Table S1), and as a result the relationship between selection and prey diversity was not significantly modified (Fig. 5B). The HMM characterized the states mostly on movement speed (i.e., step length, Fig. S1), and the states had contrasting relationships with prey diversity (Fig. 5C). The probability of being in the slow behaviour, which is often of higher importance for conservation because it is thought to be associated with foraging ([112], but see [87] for a discussion on the complexities of interpretation), had a positive relationship with prey diversity (Fig. 5C). The probability of being in the moderate speed behaviour decreased with prey diversity, while the probability of being in fast behaviour remained relatively constant with prey diversity, with a potential peak in low-medium prey diversity (Fig. 5C).

If these results were being used for conservation, the HMM would suggest that high prey diversity is important for ringed seal space use and important behaviours, while the RSF on the thinned dataset and the SSFs would suggest that this covariate is not important, likely due to the conceptual attributes of each model. For example, the different selection coefficient estimates between the RSF and the SSFs may be attributed to their distinct definition of available habitat. The RSF assumes the entire study area (within the minimum convex polygon (MCP) in this case study) is accessible to the animal, whereas the SSF generates control points at each step and availability is estimated simultaneously with habitat selection through the movement kernel, rather than being assumed a priori (Fig. 4A-B). Further, it is expected that RSFs will have greater coefficients than SSFs due to scale [113,114]. It is likely that the positive relationship with prey diversity for the RSF on the full dataset is significant due to autocorrelation in the dataset rather than the scale at which the model assesses habitat-environment relationships. Indeed, once the dataset was thinned, the estimate – while still positive and with a similar effect size – was not significant (Table S1, Fig. 5A), highlighting the importance of considering autocorrelation in the dataset (Box. 1). The RSF on the full dataset and HMM slow movement state likely show similar results because slow movement is associated with greater residency (i.e., remaining in an area) and thus will result in more locations in an area.

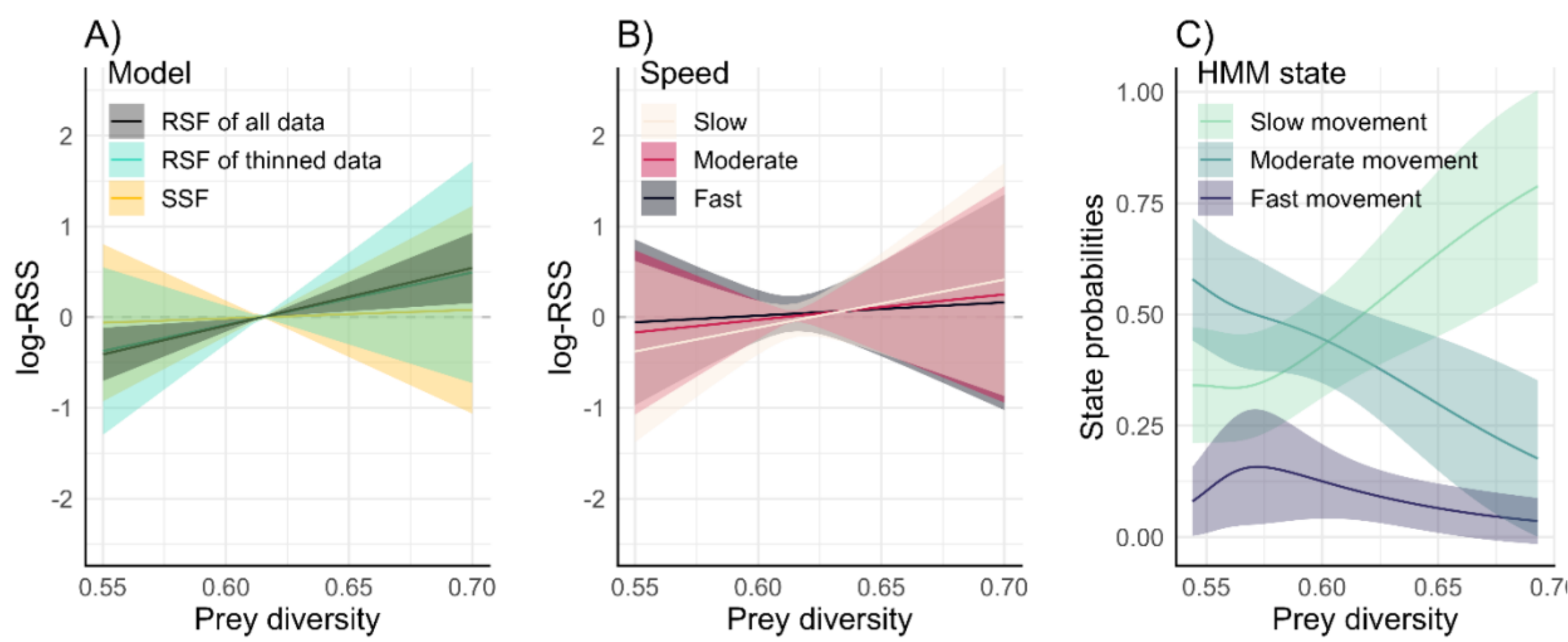

**Figure 5**. Log relative selection strength (log-RSS) from the A) RSF on the full dataset, RSF on the thinned dataset (every 10$^{th}$ location), and SSF without movement-habitat interactions and B) SSF with movement-habitat interactions, where prey diversity affected step length; log-RSS was calculated for speeds at the 25th (slow), 50th (moderate), and 75th (fast) percentiles of step length, which we infer as "speed" since the locations are recorded at a fixed time interval (see [30,111]). C) Estimated state probabilities from the HMM as a function of prey diversity (on the transition probability). Shaded areas represent the standard error. Note the bowtie shape of the standard error in A-B is due to log-RSS calculating selection strength *relative* to a starting step, in this case, where prey diversity is set to its mean.

Differences in coefficient estimates will affect maps of predicted probability of seal occurrence or behaviour (selection function or HMM, respectively; Fig. 6). Given the model's positive coefficient estimate with prey diversity, both RSFs predicted that seals are most often found in a small portion of the study area where there is high prey diversity (Fig. 6A shows the RSF on the thinned dataset; the RSF of the full dataset renders a similar map as the coefficient is similar, Table S1). The SSFs did not have significant relationships with prey diversity, thus, the maps generated from the SSUDs showed that probability of use was more uniformly distributed in space (Fig. 6B-C). The HMM will produce one map per state (behaviour), and here with vastly contrasting spatial patterns. The slow movement state, which was positively related to prey diversity, showed a similar restricted spatial pattern to the RSF (Fig. 6D), whereas the moderate movement state, which had a negative relationship with prey diversity, showed the opposite pattern to the RSF prediction map (Fig. 6E). The fast movement state did not show much spatial variation (Fig. 6F), reflective of its non-significant relationship with prey diversity (Fig. 5F). The SSF-generated maps did not show any similarities with the other models' maps (or relative to prey diversity), likely due to prey diversity not being relevant at the step-scale. Overall, we see that prey diversity may be relevant for the behavioural state transition at the whole track-level scale (HMM). We also note important considerations for the autocorrelation in the data, as once thinned, the RSF no longer estimated a significant relationship for prey diversity. This case study demonstrates how choosing the wrong model for the question at hand can mislead ecological insight and could misinform conservation and management decisions.

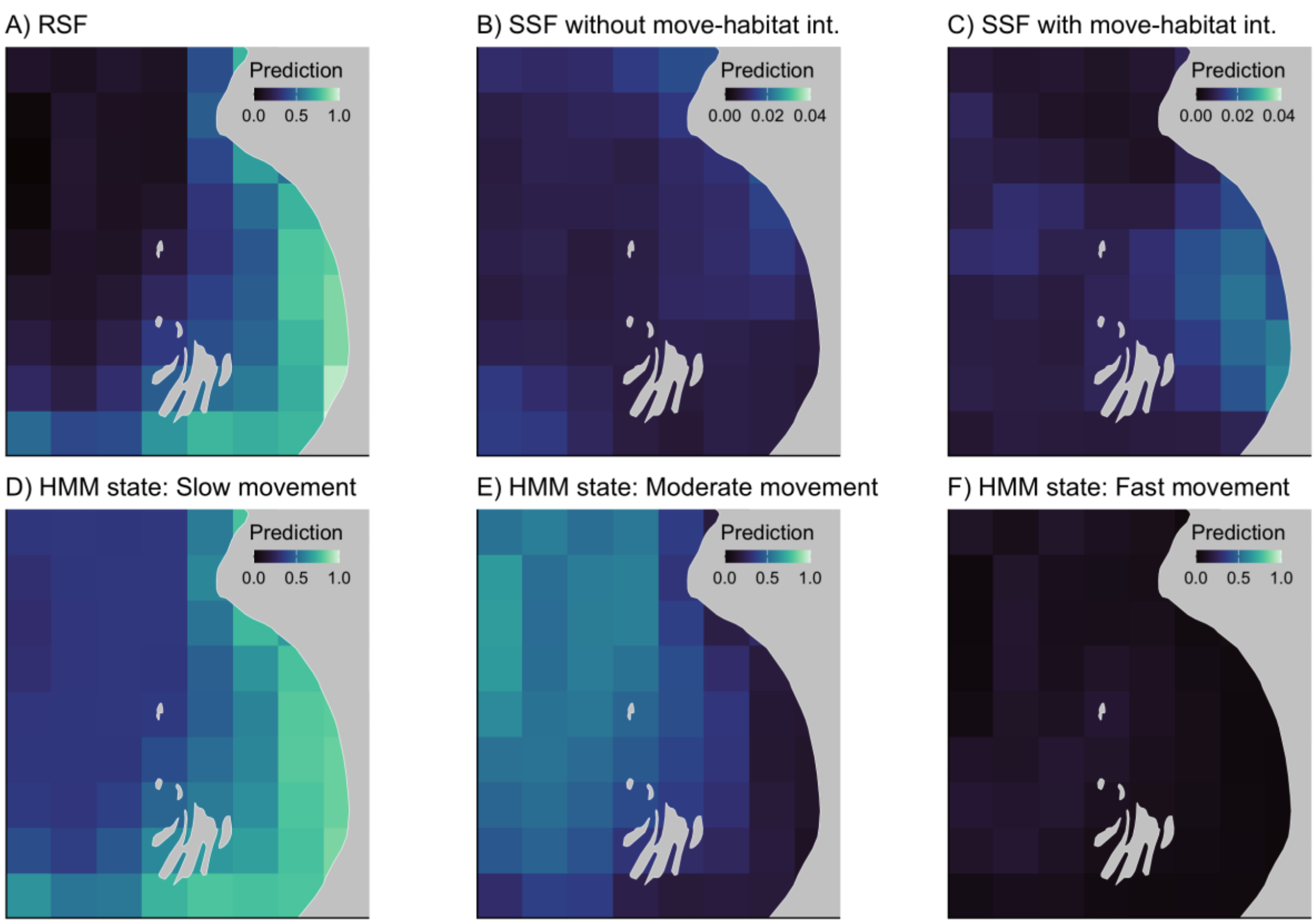

**Figure 6**. Predicted density (A-C) and predicted state probability (D-F; herein *prediction*) from A) resource selection function (RSF; thinned dataset), B) step selection function (SSF) without movement-habitat interactions, C) SSF with movement-habitat interactions, and hidden Markov

model (HMM) D) state 1: slow movement, E) state 2: moderate movement, and F) state 3: fast movement. Note that utilization distributions in B and C were created using simulated paths with 200,000 locations.

In this case study, selection at the home range scale modeling approach (i.e., RSF) may be most relevant due to the coarse resolution of the prey data. To simplify comparisons across models, this case study was limited to one seal, one prey metric, and used the default settings for these models. Thus, to reflect realistic ecological dynamics, an analysis of a more comprehensive ringed seal dataset (e.g., more individuals, longer tracking, and more prey metrics) should be performed in the future, and further, the uncertainty associated with Argos data and in generating prediction maps should be accounted for.

## 5. Future directions

Selection functions and HMMs are widely applied models that provide essential insight from animal movement data, and several new and extended versions of these models offer deeper ecological understanding. For example, recent models have incorporated memory [115], diel variation [116], and habitat connectivity [117] to better understand animal movement. Further, integration of habitat availability into HMMs (termed HMM-SSF) has been developed to provide insight on behaviour-specific habitat selection (e.g., [118–120]). Additionally, new approaches incorporate additional data streams, such as those from spatial surveys [121]. These advancements extend conventional animal-habitat models for movement data and exemplify the recent exciting progress in the field.

Recent work has also addressed limitations of mainstream selection functions and HMMs by improving model fitting, checking, and simulation capabilities. For example, models have incorporated more complex spatiotemporal structures [122–124], non-linear and random effects [96], and cross-correlated movement kernels [90]. Notably, to support the growing complexity of these models, the recent development of the `hmmTMB` package enhances the implementation of HMMs by allowing for more complex models and inclusion of random effects [125]. Additionally, new methods and tools have been introduced to evaluate model goodness-of-fit [126], address challenges associated with incomplete data [127], and ease simulation for space use predictions from SSFs [83,84].

There are a variety of machine learning approaches (e.g., MaxEnt, random forests) that we have not covered because our review is focused on statistical methods. However, it is important to note that MaxEnt is considered equivalent to an exponential RSF [46,128], and generally may be better suited for occupancy data (i.e., a location can only be used once, as in species distribution data, [129]), but can also be applied to movement data [31]. Additionally, random forests may be used for modeling animal movement and generally relax the assumptions associated with parametric methods, but the results may be more challenging to interpret [130]. If prediction is the goal, machine learning approaches including MaxEnt and random forests, or a parametric approach using least absolute shrinkage and selection operator (LASSO) for model selection, are often recommended [43,101].

## 6. Conclusions

The models reviewed in this paper for linking animals' movement data to the environment are widely applicable and can be easily implemented but provide different ecological insights. Our review highlights the conceptual differences that will aid ecologists in choosing a model based on their goals. For example, RSFs may be well suited for identifying broad corridors or protected areas, SSFs for understanding movement patterns, and HMM for understanding behaviours. Particularly, we highlight that RSFs are the most broad-scale model and are useful for home-range studies, whereas SSFs and ultimately HMMs get increasingly more movement-specific. Also, we show how RSF results change when autocorrelation is removed. Further, the sampling – or exclusion as in the case of HMMs – of availability locations can influence the coefficient estimates, as can the inclusion of movement information in the model. All these models are invaluable for ecological research, yet careful consideration of the mathematical and conceptual underpinnings is necessary to address the ecological questions and applied conservation management that these results may inform.

**List of Abbreviations**

EM – expectation-maximization

EPBC – Environment Protection and Biodiversity Conservation

ESA – Endangered Species Act

HMM – hidden Markov model

ICAR – intrinsic conditional autoregressive

INLA – integrated nested Laplace approximation

IPP – inhomogeneous Poisson point process

LASSO – least absolute shrinkage and selection operator

Log-RSS – log of the relative selection strength

MCP – minimum convex polygon

RSF – resource selection function

SARA – Species at Risk Act

SSF – step selection function

SSUD – steady state utilization distribution

## Declarations

**Ethics approval and consent to participate:** All applicable institutional and/or national guidelines for the care and use of animals were followed. Animal handling was approved by the Freshwater Institute Animal Care Committee (FWI-ACC-2012) and Fisheries and Oceans Canada's Licenses to Fish for Scientific Purposes (S-05/06-09/13-1006-NU).

**Consent for publication:** Not applicable.

**Availability of data and materials:** All data and code used in the case study are included as Appendix 3 and 4, respectively, and also available on Github (https://github.com/kflorko/movementstats_review)

**Competing interests:** The authors declare that they have no competing interests.

**Funding:** We acknowledge the financial support from Fisheries and Oceans Canada, Nunavut Wildlife Management Board, ArcticNet, Natural Sciences and Engineering Research Council of Canada (NSERC), Ocean Leaders Graduate Fellowship, Canada Research Chairs program, Canada Foundation for Innovation, B.C. Knowledge Development Fund, Weston Family Foundation,

Northern Scientific Training Program, Polar Knowledge Canada, Earth Rangers, Social Sciences and Humanities Research Council.

**Author's contributions:** KRNF, RRT, and MAM designed the study. KRNF did the analysis and prepared figures and tables. KRNF wrote the manuscript with contributions from RRT, RG, ES, and MAM. KRNF wrote the code and Github tutorial with contributions from RRT and MAM. DJY and SHF collected the seal data. All authors contributed ideas and revised the manuscript.

**Acknowledgements:** We gratefully acknowledge the Sanikiluaq hunters, the Sanikiluaq Hunters and Trappers Association, and especially Lucassie and Johnassie Ippak and their families for assistance in the field for the seal data used in the case study. We also thank the anonymous reviewers for their invaluable feedback. Their thoughtful and detailed comments significantly improved the quality of our manuscript and we appreciate the time and effort they dedicated to the review process.

## References

1. Andrewartha HG, Browning TO. An analysis of the idea of "resources" in animal ecology. J Theor Biol. 1961;1:83–97.
2. Huey RB. Physiological consequences of habitat selection. Am Nat. 1991;137:S91–115.
3. Matthiopoulos J, Fieberg J, Aarts G, Beyer HL, Morales JM, Haydon DT. Establishing the link between habitat selection and animal population dynamics. Ecol Monogr. 2015;85:413–36.
4. Hussey NE, Kessel ST, Aarestrup K, Cooke SJ, Cowley PD, Fisk AT, et al. Aquatic animal telemetry: a panoramic window into the underwater world. Science (1979). 2015;348:1255642.
5. Nathan R, Monk CT, Arlinghaus R, Adam T, Alós J, Assaf M, et al. Big-data approaches lead to an increased understanding of the ecology of animal movement. Science (1979). 2022;375:eabg1780.

6. Valeix M, Fritz H, Sabatier R, Murindagomo F, Cumming D, Duncan P. Elephant-induced structural changes in the vegetation and habitat selection by large herbivores in an African savanna. Biol Conserv. 2011;144:902–12.

7. Sells SN, Costello CM, Lukacs PM, Roberts LL, Vinks MA. Grizzly bear habitat selection across the Northern Continental Divide Ecosystem. Biol Conserv. 2022;276:109813.

8. Godvik IMR, Loe LE, Vik JO, Veiberg V, Langvatn R, Mysterud A. Temporal scales, trade-offs, and functional responses in red deer habitat selection. Ecology. 2009;90:699–710.

9. Morales JM, Moorcroft PR, Matthiopoulos J, Frair JL, Kie JG, Powell RA, et al. Building the bridge between animal movement and population dynamics. Philosophical Transactions of the Royal Society B: Biological Sciences. 2010;365:2289–301.

10. Matthiopoulos J, Fieberg JR, Aarts G. Species-Habitat Associations: Spatial data, predictive models, and ecological insights. 2023;

11. Hooten MB, Johnson DS, McClintock BT, Morales JM. Animal movement: statistical models for telemetry data. CRC press; 2017.

12. Signer J, Fieberg J, Avgar T. Animal movement tools (amt): R package for managing tracking data and conducting habitat selection analyses. Ecol Evol. 2019;9:880–90.

13. McClintock BT, Michelot T. momentuHMM: R package for generalized hidden Markov models of animal movement. Methods Ecol Evol. 2018;9:1518–30.

14. Joo R, Boone ME, Clay TA, Patrick SC, Clusella-Trullas S, Basille M. Navigating through the R packages for movement. Journal of Animal Ecology. 2020;89:248–67.

15. Alston JM, Joyce MJ, Merkle JA, Moen RA. Temperature shapes movement and habitat selection by a heat-sensitive ungulate. Landsc Ecol. 2020;35:1961–73.

16. Chimienti M, van Beest FM, Beumer LT, Desforges JP, Hansen LH, Stelvig M, et al. Quantifying behavior and life-history events of an Arctic ungulate from year-long continuous accelerometer data. Ecosphere. 2021;12.

17. Owen M. Some factors affecting food intake and selection in white-fronted geese. J Anim Ecol. 1972;79–92.

18. Johnson DH. The comparison of usage and availability measurements for evaluating resource preference. Ecology. 1980;61:65–71.

19. Nams VO. Sampling animal movement paths causes turn autocorrelation. Acta Biotheor. 2013;61:269–84.

20. Hooten MB, Lu X, Garlick MJ, Powell JA. Animal movement models with mechanistic selection functions. Spat Stat. 2020;37:100406.

21. Northrup JM, Vander Wal E, Bonar M, Fieberg J, Laforge MP, Leclerc M, et al. Conceptual and methodological advances in habitat-selection modeling: guidelines for ecology and evolution. Ecological Applications. 2022;32:e02470.

22. Mercker M, Schwemmer P, Peschko V, Enners L, Garthe S. Analysis of local habitat selection and large-scale attraction/avoidance based on animal tracking data: is there a single best method? Mov Ecol. 2021;9:20.

23. Northrup JM, Hooten MB, Anderson Jr CR, Wittemyer G. Practical guidance on characterizing availability in resource selection functions under a use–availability design. Ecology. 2013;94:1456–63.

24. Morris DW. Toward an ecological synthesis: a case for habitat selection. Oecologia. 2003;136:1–13.

25. Wiens JA. Resource systems, populations and communities. In: A new Ecology: Novel Approaches to Interactive Systems. Price P, Slobodchikoff C, Gaud W, Eds, editors. John Wiley and Sons, New York; 1984.

26. Hall LS, Krausman PR, Morrison ML. The habitat concept and a plea for standard terminology. Wildl Soc Bull. 1997;173–82.

27. Manly BFL, McDonald L, Thomas DL. Resource Selection by Animals: Statistical Design and Analysis for Field Studies. London: Chapman & Hall; 1993.

28. Boyce MS, McDonald LL. Relating populations to habitats using resource selection functions. Trends Ecol Evol. 1999;14:268–72.

29. Johnson CJ, Nielsen SE, Merrill EH, McDONALD TL, Boyce MS. Resource selection functions based on use-availability data: theoretical motivation and evaluation methods. J Wildl Manage. 2006;70:347–57.

30. Fieberg J, Signer J, Smith B, Avgar T. A ‘How to’ guide for interpreting parameters in habitat-selection analyses. Journal of Animal Ecology. 2021;90:1027–43.

31. Lele SR, Merrill EH, Keim J, Boyce MS. Selection, use, choice and occupancy: clarifying concepts in resource selection studies. Journal of Animal Ecology. 2013;82:1183–91.

32. Florko KRN, Ross TR, Ferguson SH, Northrup JM, Obbard ME, Thiemann GW, et al. The dynamic interaction between predator and prey drives mesopredator movement and foraging ecology. bioRxiv. 2023;2023.04.27.538582.

33. Chetkiewicz C-LB, Boyce MS. Use of resource selection functions to identify conservation corridors. Journal of Applied Ecology. 2009;1036–47.

34. McClure KM, Bastille-Rousseau G, Davis AJ, Stengel CA, Nelson KM, Chipman RB, et al. Accounting for animal movement improves vaccination strategies against wildlife disease in heterogeneous landscapes. Ecological Applications. 2022;32:e2568.

35. Martin ME, Moriarty KM, Pauli JN. Landscape seasonality influences the resource selection of a snow-adapted forest carnivore, the Pacific marten. Landsc Ecol. 2021;36:1055–69.

36. Knopff AA, Knopff KH, Boyce MS, Clair CCS. Flexible habitat selection by cougars in response to anthropogenic development. Biol Conserv. 2014;178:136–45.

37. Darlington S, Ladle A, Burton AC, Volpe JP, Fisher JT. Cumulative effects of human footprint, natural features and predation risk best predict seasonal resource selection by white-tailed deer. Sci Rep. 2022;12:1072.

38. Lele SR, Keim JL. Weighted distributions and estimation of resource selection probability functions. Ecology. 2006;87:3021–8.

39. Fithian W, Hastie T. Statistical models for presence-only data: finite-sample equivalence and addressing observer bias. Ann Appl Stat. 2012;

40. Hastie T, Fithian W. Inference from presence-only data; the ongoing controversy. Ecography. 2013;36:864–7.

41. Johnson DS, Hooten MB, Kuhn CE. Estimating animal resource selection from telemetry data using point process models. Journal of Animal Ecology. 2013;82:1155–64.

42. Warton DI, Shepherd LC. Poisson point process models solve the" pseudo-absence problem" for presence-only data in ecology. Ann Appl Stat. 2010;1383–402.

43. Elith J, Leathwick JR. Species distribution models: ecological explanation and prediction across space and time. Annu Rev Ecol Evol Syst. 2009;40:677–97.

44. Thurfjell H, Ciuti S, Boyce MS. Applications of step-selection functions in ecology and conservation. Mov Ecol. 2014;2:1–12.

45. Michelot T, Klappstein NJ, Potts JR, Fieberg J. Understanding step selection analysis through numerical integration. Methods Ecol Evol. 2024;1:24–35.
46. Aarts G, Fieberg J, Matthiopoulos J. Comparative interpretation of count, presence–absence and point methods for species distribution models. Methods Ecol Evol. 2012;3:177–87.
47. Hebblewhite M, Merrill E. Modelling wildlife–human relationships for social species with mixed-effects resource selection models. Journal of applied ecology. 2008;45:834–44.
48. Muff S, Signer J, Fieberg J. Accounting for individual-specific variation in habitat-selection studies: Efficient estimation of mixed-effects models using Bayesian or frequentist computation. Journal of Animal Ecology. 2020;89:80–92.
49. Boyce MS. Scale for resource selection functions. Divers Distrib. 2006;12:269–76.
50. DeCesare NJ, Hebblewhite M, Schmiegelow F, Hervieux D, McDermid GJ, Neufeld L, et al. Transcending scale dependence in identifying habitat with resource selection functions. Ecological Applications. 2012;22:1068–83.
51. Nielsen SE, Boyce MS, Stenhouse GB, Munro RHM. Modeling grizzly bear habitats in the Yellowhead ecosystem of Alberta: taking autocorrelation seriously. Ursus. 2002;45–56.
52. Hooten MB, Hanks EM, Johnson DS, Alldredge MW. Reconciling resource utilization and resource selection functions. Journal of Animal Ecology. 2013;82:1146–54.
53. Fieberg J, Matthiopoulos J, Hebblewhite M, Boyce MS, Frair JL. Correlation and studies of habitat selection: problem, red herring or opportunity? Philosophical Transactions of the Royal Society B: Biological Sciences. 2010;365:2233–44.
54. Alston JM, Fleming CH, Kays R, Streicher JP, Downs CT, Ramesh T, et al. Mitigating pseudoreplication and bias in resource selection functions with autocorrelation-informed weighting. Methods Ecol Evol. 2023;14:643–54.
55. Koper N, Manseau M. A guide to developing resource selection functions from telemetry data using generalized estimating equations and generalized linear mixed models. Rangifer. 2012;32:195–204.
56. Avgar T, Potts JR, Lewis MA, Boyce MS. Integrated step selection analysis: bridging the gap between resource selection and animal movement. Methods Ecol Evol. 2016;7:619–30.
57. Forester JD, Im HK, Rathouz PJ. Accounting for animal movement in estimation of resource selection functions: sampling and data analysis. Ecology. 2009;90:3554–65.

58. Fortin D, Beyer HL, Boyce MS, Smith DW, Duchesne T, Mao JS. Wolves influence elk movements: Behavior shapes a trophic cascade in Yellowstone National Park. Ecology [Internet]. 2005;86:1320–30. Available from: http://www.esajournals.org/doi/abs/10.1890/04-0953%5Cn%3CGo to ISI%3E://000228960000027

59. Rhodes JR, McAlpine CA, Lunney D, Possingham HP. A spatially explicit habitat selection model incorporating home range behavior. Ecology. 2005;86:1199–205.

60. Prima M-C, Duchesne T, Fortin D. Robust inference from conditional logistic regression applied to movement and habitat selection analysis. PLoS One. 2017;12:e0169779.

61. Craiu R V, Duchesne T, Fortin D. Inference methods for the conditional logistic regression model with longitudinal data. Biometrical Journal: Journal of Mathematical Methods in Biosciences. 2008;50:97–109.

62. Squires JR, DeCesare NJ, Olson LE, Kolbe JA, Hebblewhite M, Parks SA. Combining resource selection and movement behavior to predict corridors for Canada lynx at their southern range periphery. Biol Conserv. 2013;157:187–95.

63. Roever CL, Boyce MS, Stenhouse GB. Grizzly bear movements relative to roads: application of step selection functions. Ecography. 2010;33:1113–22.

64. Webber QMR, Prokopenko CM, Kingdon KA, Turner JW, Vander Wal E. Effects of the social environment on movement-integrated habitat selection. Mov Ecol. 2024;1:61.

65. Eisaguirre JM, Booms TL, Barger CP, Lewis SB, Breed GA. Novel step selection analyses on energy landscapes reveal how linear features alter migrations of soaring birds. Journal of Animal Ecology. 2020;89:2567–83.

66. Patterson TA, Basson M, Bravington M V, Gunn JS. Classifying movement behaviour in relation to environmental conditions using hidden Markov models. Journal of Animal Ecology. 2009;78:1113–23.

67. Glennie R, Adam T, Leos-Barajas V, Michelot T, Photopoulou T, McClintock BT. Hidden Markov models: Pitfalls and opportunities in ecology. Methods Ecol Evol. 2023;14:43–56.

68. Zucchini W, MacDonald IL. Hidden Markov models for time series: an introduction using R. Chapman and Hall/CRC; 2009.

69. McClintock BT, Langrock R, Gimenez O, Cam E, Borchers DL, Glennie R, et al. Uncovering ecological state dynamics with hidden Markov models. Ecol Lett. 2020;23:1878–903.

70. Gardiner R, Hamer R, Leos-Barajas V, Peñaherrera-Palma C, Jones ME, Johnson C. State-space modeling reveals habitat perception of a small terrestrial mammal in a fragmented landscape. Ecol Evol. 2019;9:9804–14.

71. Togunov RR, Derocher AE, Lunn NJ, Auger-Méthé M. Drivers of polar bear behavior and the possible effects of prey availability on foraging strategy. Mov Ecol. 2022;10:50.

72. Creel S, Merkle J, Mweetwa T, Becker MS, Mwape H, Simpamba T, et al. Hidden Markov Models reveal a clear human footprint on the movements of highly mobile African wild dogs. Sci Rep. 2020;10:17908.

73. Morales JM, Haydon DT, Frair J, Holsinger KE, Fryxell JM. Extracting more out of relocation data: building movement models as mixtures of random walks. Ecology. 2004;85:2436–45.

74. Langrock R, King R, Matthiopoulos J, Thomas L, Fortin D, Morales JM. Flexible and practical modeling of animal telemetry data: hidden Markov models and extensions. Ecology. 2012;93:2336–42.

75. Ngô MC, Heide-Jørgensen MP, Ditlevsen S. Understanding narwhal diving behaviour using Hidden Markov Models with dependent state distributions and long range dependence. PLoS Comput Biol [Internet]. 2019;15:e1006425. Available from: https://doi.org/10.1371/journal.pcbi.1006425

76. Michelot T, Langrock R, Patterson TA. moveHMM: an R package for the statistical modelling of animal movement data using hidden Markov models. Methods Ecol Evol. 2016;7:1308–15.

77. Auger-Méthé M, Newman K, Cole D, Empacher F, Gryba R, King AA, et al. A guide to state–space modeling of ecological time series. Ecol Monogr. 2021;91:e01470.

78. Breed GA, Costa DP, Jonsen ID, Robinson PW, Mills-Flemming J. State-space methods for more completely capturing behavioral dynamics from animal tracks. Ecol Modell. 2012;235:49–58.

79. Gryba RD, Wiese FK, Kelly BP, Von Duyke AL, Pickart RS, Stockwell DA. Inferring foraging locations and water masses preferred by spotted seals Phoca largha and bearded seals Erignathus barbatus. Mar Ecol Prog Ser. 2019;631:209–24.

80. McClintock BT, Johnson DS, Hooten MB, Ver Hoef JM, Morales JM. When to be discrete: the importance of time formulation in understanding animal movement. Mov Ecol. 2014;2:1–14.

81. Sidrow E, Heckman N, Fortune SME, Trites AW, Murphy I, Auger-Méthé M. Modelling multi-scale, state-switching functional data with hidden Markov models. Canadian Journal of Statistics. 2022;50:327–56.

82. Signer J, Fieberg J, Avgar T. Estimating utilization distributions from fitted step-selection functions. Ecosphere. 2017;8:e01771.

83. Signer J, Fieberg J, Reineking B, Schlägel U, Smith B, Balkenhol N, et al. Simulating animal space use from fitted integrated Step-Selection Functions (iSSF). Methods Ecol Evol. 2024;15:43–50.

84. Potts JR, Börger L. How to scale up from animal movement decisions to spatiotemporal patterns: An approach via step selection. Journal of Animal Ecology. 2023;92:16–29.

85. Buderman FE, Helm PJ, Clark JD, Williamson RH, Yarkovich J, Mullinax JM. A multi-level modeling approach to guide management of female feral hogs in Great Smoky Mountains National Park. Biol Invasions. 2023;25:3065–82.

86. Buderman FE, Gingery TM, Diefenbach DR, Gigliotti LC, Begley-Miller D, McDill MM, et al. Caution is warranted when using animal space-use and movement to infer behavioral states. Mov Ecol. 2021;9:1–12.

87. Florko KRN, Shuert CR, Cheung WWL, Ferguson SH, Jonsen ID, Rosen DAS, et al. Linking movement and dive data to prey distribution models: new insights in foraging behaviour and potential pitfalls of movement analyses. Mov Ecol. 2023;11:17.

88. Boyce MS, Vernier PR, Nielsen SE, Schmiegelow FKA. Evaluating resource selection functions. Ecol Modell. 2002;157:281–300.

89. Thompson PR, Derocher AE, Edwards MA, Lewis MA. Detecting seasonal episodic-like spatio-temporal memory patterns using animal movement modelling. Methods Ecol Evol. 2022;13:105–20.

90. Hodel FH, Fieberg JR. Circular–linear copulae for animal movement data. Methods Ecol Evol. 2022;13:1001–13.

91. Arce Guillen R, Lindgren F, Muff S, Glass TW, Breed GA, Schlägel UE. Accounting for unobserved spatial variation in step selection analyses of animal movement via spatial random effects. Methods Ecol Evol. 2023;14:2639–53.

92. Forrest SW, Pagendam D, Bode M, Drovandi C, Potts JR, Perry J, et al. Predicting fine-scale distributions and emergent spatiotemporal patterns from temporally dynamic step selection simulations. Ecography. 2024;e07421.

93. Dejeante R, Valeix M, Chamaillé-Jammes S. Time-varying habitat selection analysis: A model and applications for studying diel, seasonal, and post-release changes. Ecology. 2024;105:e4233.

94. Arce Guillen R, Pohle J, Jeltsch F, Roeleke M, Reineking B, Klappstein N, et al. Flexible movement kernel estimation in habitat selection analyses with generalized additive models. bioRxiv. 2024;2024–6.

95. Hodel FH, Fieberg JR. Circular–linear copulae for animal movement data. Methods Ecol Evol. 2022;13:1001–13.

96. Klappstein NJ, Michelot T, Fieberg J, Pedersen EJ, Mills Flemming J. Step selection functions with non-linear and random effects. Methods Ecol Evol. 2024;15:1332–46.

97. Warton DI, Blanchet FG, O'Hara RB, Ovaskainen O, Taskinen S, Walker SC, et al. So many variables: joint modeling in community ecology. Trends Ecol Evol. 2015;30:766–79.

98. Potts JR, Börger L, Strickland BK, Street GM. Assessing the predictive power of step selection functions: How social and environmental interactions affect animal space use. Methods Ecol Evol. 2022;13:1805–18.

99. Langrock R, Hopcraft JGC, Blackwell PG, Goodall V, King R, Niu M, et al. Modelling group dynamic animal movement. Methods Ecol Evol. 2014;5:190–9.

100. McLoughlin PD, Morris DW, Fortin D, Vander Wal E, Contasti AL. Considering ecological dynamics in resource selection functions. Journal of animal ecology. 2010;79:4–12.

101. Gerber BD, Northrup JM. Improving spatial predictions of animal resource selection to guide conservation decision making. Ecology. 2020;101:e02953.

102. Matthiopoulos J, Hebblewhite M, Aarts G, Fieberg J. Generalized functional responses for species distributions. Ecology. 2011;92:583–9.

103. Brost BM, Hooten MB, Hanks EM, Small RJ. Animal movement constraints improve resource selection inference in the presence of telemetry error. Ecology. 2015;

104. Lawler E, Whoriskey K, Aeberhard WH, Field C, Mills Flemming J. The conditionally autoregressive hidden Markov model (carhmm): Inferring behavioural states from animal tracking data exhibiting conditional autocorrelation. J Agric Biol Environ Stat. 2019;24:651–68.

105. Florko KRN, Tai TC, Cheung WWL, Sumaila UR, Ferguson SH, Yurkowski DJ, et al. Predicting how climate change threatens the prey base of Arctic marine predators. Dryad. 2021;dataset.

106. Florko KRN, Tai TC, Cheung WWL, Ferguson SH, Sumaila UR, Yurkowski DJ, et al. Predicting how climate change threatens the prey base of Arctic marine predators. Ecol Lett. 2021;24:2563–75.

107. Ferguson SH, Young BG, Yurkowski DJ, Anderson R, Willing C, Nielsen O. Demographic, ecological and physiological responses of ringed seals to an abrupt decline in sea ice availability. PeerJ. 2017;

108. Jonsen ID, Grecian WJ, Phillips L, Carroll G, McMahon C, Harcourt RG, et al. aniMotum, an R package for animal movement data: Rapid quality control, behavioural estimation and simulation. Methods Ecol Evol. 2023;14:806–16.

109. Costa DP, Robinson PW, Arnould JPY, Harrison A, Samantha E, Hassrick JL, et al. Accuracy of ARGOS locations of pinnipeds at-sea estimated using Fastloc GPS. PLoS One. 2010;5.

110. Forester JD, Ives AR, Turner MG, Anderson DP, Fortin D, Beyer HL, et al. State–space models link elk movement patterns to landscape characteristics in Yellowstone National Park. Ecol Monogr. 2007;77:285–99.

111. Avgar T, Lele SR, Keim JL, Boyce MS. Relative selection strength: Quantifying effect size in habitat-and step-selection inference. Ecol Evol. 2017;7:5322–30.

112. Pyke GH, Pulliam HR, Charnov EL. Optimal foraging: a selective review of theory and tests. Q Rev Biol. 1977;52:137–54.

113. Moorcroft PR, Barnett A. Mechanistic home range models and resource selection analysis: a reconciliation and unification. Ecology. 2008;89:1112–9.

114. Schlägel UE, Lewis MA. Robustness of movement models: can models bridge the gap between temporal scales of data sets and behavioural processes? J Math Biol. 2016;73:1691–726.

115. Kim D, Thompson PR, Wolfson DW, Merkle JA, Oliveira-Santos LGR, Forester JD, et al. Identifying signals of memory from observations of animal movements. Mov Ecol. 2024;12:72.

116. Feldmann CC, Mews S, Coculla A, Stanewsky R, Langrock R. Flexible modelling of diel and other periodic variation in hidden Markov models. J Stat Theory Pract. 2023;17:45.

117. Valle D, Attias N, Cullen JA, Hooten MB, Giroux A, Oliveira-Santos LGR, et al. Bridging the gap between movement data and connectivity analysis using the Time-Explicit Habitat Selection (TEHS) model. Mov Ecol. 2024;12:19.

118. Nicosia A, Duchesne T, Rivest L-P, Fortin D. A multi-state conditional logistic regression model for the analysis of animal movement. 2017;

119. Prima M-C, Duchesne T, Merkle JA, Chamaillé-Jammes S, Fortin D. Multi-mode movement decisions across widely ranging behavioral processes. PLoS One. 2022;17:e0272538.

120. Klappstein NJ, Thomas L, Michelot T. Flexible hidden Markov models for behaviour-dependent habitat selection. Mov Ecol. 2023;11:30.

121. Blackwell PG, Matthiopoulos J. Joint inference for telemetry and spatial survey data. Ecology. 2024;e4457.

122. Eisaguirre JM, Williams PJ, Hooten MB. Rayleigh step-selection functions and connections to continuous-time mechanistic movement models. Mov Ecol. 2024;12:14.

123. Michelot T, Gloaguen P, Blackwell PG, Étienne M. The Langevin diffusion as a continuous-time model of animal movement and habitat selection. Methods Ecol Evol. 2019;10:1894–907.

124. Michelot T, Blackwell PG. State-switching continuous-time correlated random walks. Methods Ecol Evol. 2019;10:637–49.

125. Michelot T. hmmTMB: Hidden Markov models with flexible covariate effects in R. arXiv preprint arXiv:221114139. 2022;

126. Fieberg J, Freeman S, Signer J. Using lineups to evaluate goodness of fit of animal movement models. Methods Ecol Evol. 2024;

127. Hofmann DD, Cozzi G, Fieberg J. Methods for implementing integrated step-selection functions with incomplete data. Mov Ecol. 2024;12:37.

128. Renner IW, Warton DI. Equivalence of MAXENT and Poisson point process models for species distribution modeling in ecology. Biometrics. 2013;69:274–81.

129. Elith J, Phillips SJ, Hastie T, Dudík M, Chee YE, Yates CJ. A statistical explanation of MaxEnt for ecologists. Divers Distrib. 2011;17:43–57.
130. Wijeyakulasuriya DA, Eisenhauer EW, Shaby BA, Hanks EM. Machine learning for modeling animal movement. PLoS One. 2020;15:e0235750.